Title: Trademark filings and patent application count time series are structurally near-identical and cointegrated: Implications for studies in innovation

Author: Iraj Daizadeh, PhD, Global Regulatory Affairs, Takeda Pharmaceuticals, 40 Landsdowne Street, Cambridge, MA 02139. All correspondence should be sent to: iraj.daizadeh@takeda.com

Abstract:

Through time series analysis, this paper empirically explores, confirms and extends the trademark/patent inter-relationship as proposed in the normative intellectual-property (IP)-oriented Innovation Agenda view of the science and technology (S&T) firm. Beyond simple correlation, it is shown that trademark-filing (Trademarks) and patent-application counts (Patents) have similar (if not, identical) structural attributes (including similar distribution characteristics and seasonal variation, cross-wavelet synchronicity/coherency (short-term cross-periodicity) and structural breaks) and are cointegrated (integration order of 1 – I(1)) over a period of approximately 40 years (given the monthly observations). The existence of cointegration strongly suggests a "long-run" equilibrium between the two indices; that is, there is (are) exogenous force(s) restraining the two indices from diverging from one another. Structural breakpoints in the chrono-dynamics of the indices supports the existence of potentially similar exogeneous forces(s), as the break dates are simultaneous/near-simultaneous (Trademarks: 1987, 1993, 1999, 2005, 2011; Patents: 1988, 1994, 2000, and 2011). A discussion of potential triggers (affecting both time series) causing these breaks, and the concept of equilibrium in the context of these proxy measures are presented. The cointegration order and structural co-movements resemble other macro-economic variables, stoking the opportunity of using econometrics approaches to further analyze these data. As a corollary, this work further supports the inclusion of trademark analysis in innovation studies. Lastly, the data and corresponding analysis tools (R program) are presented as Supplementary Materials for reproducibility and convenience to conduct future work for interested readers.

Disclosures and Disclaimers:

The author is an employee of Takeda Pharmaceuticals; however, this work was completed independently of his employment. The views expressed in this article may not represent those of Takeda Pharmaceuticals.

## Introduction

One of the more common methods for inquiring about the dynamics (e.g., rates, structure) of innovativeness in science and technology (IS&T) firms is via intellectual property (IP)-related metrics (Dziallas and Blind, 2018). Simplistically, the rationale of using IP-related proxy measures of innovation primarily rests on the nature of the output (viz., inventions) generated by such firms (Daizadeh et al., 2002). Notably, sponsors may seek one or more patents to protect an invention, assuming such IP meets certain evidentiary standards of utility, novelty, and non-obviousness and perceived future economic rents justify a patent over that of publishing or retaining the knowledge as a trade-secret (ibid). Therefore, one can understand the intrinsic concept captured in a patent, and that the greater number of such IP assets implies greater innovativeness. Optimizing IP generation (and thus innovativeness) has resulted in S&T firms reorienting their organizations, systems, and processes accordingly (Daizadeh, 2003, 2007).

Conceptually, it is more challenging to extend the logic of innovativeness to other forms of IP, especially to that of trademark-related metrics (e.g., filings) of IS&T firms, as the criteria for meriting a trademark is a more amorphous entity, generally defined as 'word, phrase, symbol, and/or design that identifies and distinguishes the source of the goods of one party from those of other[1].' Some researchers have expressed significant concern over the use of trademarks of IS&T firms. For example, Hipp and Grupp (2005) state "even services containing no or only low levels of innovation can be brand protected. This limits the trademarks statistics' value as an innovation indicator (ibid, p 526)." Others have been more nuanced with their criticism of the approach, considering the topic as one of definition (Flikkema et al,

[1] https://www.uspto.gov/trademarks-getting-started/trademark-basics/trademark-patent-or-copyright viewed on 19-Nov-2019.

2015), and are leveraging relatively recent research viewing the link between trademarks and innovation accordingly. For exhaustive accounts on the literature on such topics, readers are referred to Dziallas and Blind (2019), Siekierski, et al., (2018), among others.

In the past, among other interesting metrics, Daizadeh (2007, 2009) found strong (>96%; p-value < 0.0001) correlation between patent applications (Patents) and trademarks filings (Trademarks) over a multi-year (1970-2002) period. In the same work, Daizadeh also proposed that the normative model, termed herein the 'Daizadeh Innovation Agenda (DIA)' (see Figure 1 in Daizadeh, 2009), which was conjectured from the legal and financial nature of patents, trademarks, publications, and press releases, may help in understanding the identified correlative patterns. The Innovation Agenda, as proposed, is to be comprised of several socio-economic S&T firm-specific metrics including: US Research and Development Spend, Number of Trademark Filings and Registrations, Number of Patent Applications and Issuances, Number of Press Releases, and various S&T related Stock Indices. While the data used (as is here) is US specific, it may be generalizable to any intellectual-property based entrepreneurial S&T intensive jurisdiction. In that work, Daizadeh found – using annual values for these metrics and based on correlation and partial-correlation analysis – a series of interesting findings including correlation and partial correlation across these selected metrics.

Lastly, the use of Trademark data in innovation studies has been of recent interest. Daizadeh (2007, 2009) added empirical support for the conjecture that if Patents were a proxy measure of innovation, and if Trademarks were strongly correlated with Patents, then Trademarks may also play a role as a proxy measure of innovation. Recently there has been renewed interest in the use of Trademarks (and related data) as measures of innovation (see, e.g., citations in Dziallis and Blind, 2019). Characterizing the inter-relatedness between these metrics beyond that of correlation may further strengthen the argument that Trademarks be a standard measure of innovativeness in S&T firms.

Time series statistical analysis is prevalent across scientific disciplines and includes a diverse assortment of approaches. Statistical practices of relevance to this paper, as we seek to understand the temporal co-mobility (inter-relatedness) of Patents and Trademarks (bivariates), include descriptive statistics (e.g., seasonal variation), cointegration, structural break / change point, and cross-wavelet analyses. While several approaches may have been taken to investigate inter-relatedness of a bivariate system, the methods described below were selected due to several factors including: ease of access, ease of interpretation, prevalence of use (and thus greater confidence in strength and limitations of methods), and intrinsic criteria of data (e.g., non-normality).

Various descriptive statistics were performed to provide insight into the distribution (e.g., normality) and stability (e.g., stationarity), and to better select further statistical analyses. As previously mentioned, Daizadeh (2007, 2009) found a strong correlation between Patents and Trademarks. Using time series decomposition and cross-wavelet analyses, as a qualitative tool, the nature of the correlation (synchronicity) was further explored including elucidating periodicity contributing the most to the correlation pattern. This work empirically explores monthly observations over a period of roughly 40 years for both time series. Generally, this report finds that Trademarks and Patents (time series) are similar statistical moments (mean, variance, skew, and kurtosis), non-normal and non-stationary with seasonal variation, with short-term periodicity, across all years explored. A cross-wavelet analysis was also performed to obtain a view into latent periodicity. The analysis reconfirmed the high correlation but resolved interesting dynamics in short-term periodicity associated over the most recent decade.

Cointegration (unlike correlation) analysis captures so-called ‘long-run equilibrium’ derived from stochastic relationships restricting co-movement divergence between the timeseries under-study (Granger, 1981; Engle and Granger, 1987; Dolado, et al., 1999). That is, cointegration regards the degree of differences in the timeseries as opposed to the directionality of the co-movement (e.g., positive correlation in which the bivariates move in the same direction). The cointegration statistical analyses, included those of the Johansen and Philips and Ouliaris tests, confirmed that the time series were co-moving, implying that some exogeneous effect(s) were imposing a constraint on this system.

As other statistics, cointegration may be affected by a non-trivial change in the course of the timeseries, which may be termed a 'structural break,' 'structural change,', or a 'regime shift.' While there are several definitions for a structural break, and thus methods to elucidate or predict such changes, the following is illustrative: "Structural break as an unpredictable event in which the relationship among the variables in a model changes, and this change cannot be predicted in any sense from past data (Maheu and Gordon, 2008)." Should such abrupt changes occur (quasi)-simultaneously, then it may be presumed that the same exogeneous event affects both Trademarks and Patents, adding further (if not confirmatory) evidence of not only inter-relatedness between the variables. Here, it is found from generalized fluctuation tests that structural breaks exist and identify and date the breakpoints as: Trademarks: 1987, 1993, 1999, 2005, 2011; Patents: 1988, 1994, 2000, and 2011, using standard models (see below for details and associated citations). A discussion of potential triggers for these dates is presented below.

In this paper, assuming the DIA model for the IP-intensive S&T firm, the base hypothesis explored is that if Patents and Trademarks are both affected by the same and/or similar exogenous variables, then their respective timeseries should be 'inter-related.' Further, these data add support to an assumption proposed in the DIA; namely, that an exogeneous factor(s) was applied to patent applications and trademark filings, leading to the inter-relationship. Bivariate inter-relatedness is explored empirically using descriptive statistics, structural break point, and cointegration analyses on the bivariate monthly timeseries over an extended period (1977-2016; see Methodology) compared with the original Daizadeh paper. Observation of simultaneous/quasi-simultaneous structural changes, existence of cointegration (which would imply a 'long term equilibrium' restraining the differences between the timeseries), and other structural co-movements (such as synchronicity, coherence) in the bivariate timeseries would support the theory that common (or similar) exogeneous factors exist, and thus further add additional supportive evidence to DIA theory (necessary for formalizing further study), as well as illustrates the import of trademarks to the innovation process and thus to IS&T firms generally (see Results). This manuscript concludes with a discussion of the assumptions and limitations of the approach, and avenues for future development.

All datasets and the R Program script is presented in the Supplementary Materials section of this manuscript for reproducibility and convenience to conduct future work. Interested readers are strongly encouraged to either try their own approaches to investigate the structure of the bivariate timeseries with or without considering the materials provided.

## Methodology:

### *Data sources and preparation:*

The data were comprised of the monthly number of US patent applications (Patents) and the monthly number of US trademarks filings (Trademarks) from 1977 to 2016.

*Number of patent applications and trademark filings*: The data on Patents and Trademarks were obtained from the respective publicly available websites supported by the United States Patent and Trademark Office (USPTO) as described below:

| Variable | USPTO Publicly Available Search Site | Search Characteristics* |
|---|---|---|
| Patents | http://patft.uspto.gov/netahtml/PTO/search-adv.htm | Application Filing |
| | | Date: "APD/MM/$/YYYY" |
| Trademarks | http://tmsearch.uspto.gov/bin/gate.exe?f=tess&state=4804:57thz4.1.1 | Filing Date: "(YYYYMM$)[FD]" |
| MM/YYYY is the two/four digital representation for month/year | | |

Two searches were manually executed, resulting in 472 datapoints for each variable and captured in Excel for import into R. The 472 datapoints for each variable represents monthly observations over the period of study (approximately 40 years). The data is presented in the Supplementary Materials section of the manuscript for ease of reference and for the sake of reproducibility.

*Statistical Analysis.*

Methodology followed standard implementation, and default parameters were used throughout. While the R code (R Core Team, 2019) is presented in the Supplemental Materials section of this manuscript for reproducibility, the general algorithm for the analysis is as follows:

- Load bivariate timeseries, identify and replace outliers with average of prior and posterior-month values (R package 'tsoutliers' (López-de-Lacalle, 2019). Note: 3 outliers were determined for Trademarks (September 1982; November 1989; and June 1999) and 4 for Patents (September 1982, June 1995, October 2007, and March 2013).
- Decompose data and perform descriptive statistics, including deriving kurtosis, skew (R package 'moments' (Komsta and Novomestky, 2015)), nonparametric (Spearman and Kendall) correlation coefficients (ibid), and cross-wavelet analyses (R Package 'biwavelet,' (Gouhier, et al. 2019)) on full timeseries.
- Test for structural breakpoints (SBPs) using empirical fluctuation processes (R package 'strucchange' (Zeileis, et al., 2002, 2003)):

- Ordinary least squares-cumulative sums and moving sums (OLS-CUSUM, MOSUM)) and recursive (REC) CUSUM and REC-MOSUM (see p. 4/5 of Zeileis et al, 2002)
  - Significance testing (see p. 9 of ibid)
- Date SBPs and define segments:
  - Bai-Perron breakpoint method (Bai and Perron, 2003; R package 'strucchange' (see Zeileis, et al., 2002, 2003; see p. 112/113 in Zeileis, et al., 2003). A segment is defined as the longest length of time between two SBPs (either Patents or Trademarks).
- Test for cointegration using Johansen Procedure and Phillips and Ouliaris (Pz) tests (R package 'ucra' (Pfaff, 2008)) for full timeseries and segments.
- Determine order of integration using Kwiatkowski-Phillips-Schmidt-Shin (KPSS), Augmented Dickey–Fuller (ADF), and the Phillips and Peron tests (R package 'forecast' (Hyndman, et al., 2019; Hyndman and Khandakar, 2008) for full timeseries and segments.

<u>Results:</u>

*Descriptive Statistics:*

The distributions of the two time-series were similar (e.g., approximately symmetric (skew) and platykurtic) and thus no transformation was performed on the data (Table 1). The respective trends of the time-series generally evolve in time in an 'exponential manner,' both have similar per-annum quarterly seasonal effects, with increased contributions from the stochastic (random) elements post-2010, with a spike at circa 2000 and circa 1995 for Trademarks and Patents, respectively (Figures 1 and 2). Qualitatively it would seem that Trademarks present somewhat greater degree of 'randomness' than Patents.

< Insert Table 1, Figure 1, and Figure 2 here.

Caption table 1: Descriptive statistics of Trademarks and Patents

Caption Figure 1: Timeseries decomposition of Trademarks

Caption Figure 2: Timeseries decomposition of Patents>

*Correlation and Cross-Wavelet Analysis*

Spearman and Kendall analysis finds a strong coefficient of correlation of 0.94 and 0.80, respectively, between Trademarks and Patents; this reconfirms the work of Daizadeh (2007, 2009) for an extended period of time (1977-2016). Further examination using cross-wavelet analysis shows broadly high to very high (1) coherency (red to dark-red splotches) across the Trademark/Patent spectra and periods. Relatively low periods post-1995 and more uniformly post-2002 demonstrate increased bivariate synchronicity (at 5% statistical significance).

< Insert Figure 3 here.

Caption Figure 3: Cross-wavelet analysis of Trademarks and Patents; solid black contour lines designate the 5% significance>

*Existence, testing, and Dating of Structural Break points, and subsequent segmenting of the timeseries:*

The empirical fluctuation processes (EFPs) (ordinary least squares (OLS) and recursive modeling (REC)) test the null hypothesis of "'no structural change' [which] should be rejected when the fluctuation of the empirical processes gets improbably large compared to the fluctuation of the limiting process (Zeileis, et al., 2002, p. 6)." The EFPs were executed with a significance criterion (alpha) of 5%. The results of these four tests for Trademarks are presented in Figure 4; a similar result was found for all tests for Patents (figure not shown but calculations may be reproduced in the Supplemental Materials). Significance testing for the existence of SBPs are presented in Table 2, with p-values less than or equal to 0.01.

As can be seen in either of Figures 4 and Table 2, the EFPs cross the critical value boundary, and therefore rejecting the null hypothesis of no SBP at the 5% level. Further, the complex structures for both (Approvals and Guidances) timeseries across the tests suggest multiple structural breakpoints (Zeileis et al., 2005).

< Insert Figure 4 and Table 2 here.

Caption Figure 4: Existence of structural breakpoints for Trademarks:

Caption Table 2: Significance testing (p-value) for the existence of structural breakpoints in Trademarks and Patents>

*Dating of structural breakpoints*

The general idea of the Bai-Perron dynamic programming algorithm to date the structural breakpoints is to elucidate the breakpoints through minimizing the residual sum of squares of a linear regression model (additional details may be found in Bai and Perron, 2003 and Zeileis, et al., 2003). SBPs (including confidence limits) are presented in Figure 5 and in Table 3.

< Insert Figure 5 and Table 3 here.

Caption Figure 5: Structural breakpoints with corresponding confidence intervals (see Table 3) identified in Trademarks (black) and Patents (red)

Caption Table 3: Dating (via Bai-Perron) of the structural break points in Trademarks and Patents >

The data suggests several segments in which there is no abrupt changes in the intrinsic variability of the timeseries (stationarity). Thus, several time segments of stationarity were elucidated, and affords an ease in further analyses given minimal statistical variance/fluctuations. The existence of stationarity during these periods of time strongly suggests the lack of strength of any exogeneous factor (e.g., promulgation of novel legal frameworks and/or technologies) on the time-course of these variables. Thus, the heuristic was defined to be the longest time between Trademarks and Patents structural break points (Table 4); 6 such time-segments were identified and used for the rest of the analysis.

< Insert Table 4 here.

Caption Table 4: Segments identified as longest length of time between Trademarks and Patents structural break points (see Results for description)>

*Cointegration and the maximum order of integration (I(d)):*

Results for the Johansen Procedure and Phillips and Ouliaris tests demonstrated cointegration at alpha </= 1% for the full bivariate timeseries and the third-, fifth-, and sixth-time segments. The size of the test statistic is notable for both tests across the full timeseries (Table 5).

< Insert Table 5 here.

Caption Table 5: Results of cointegration tests across full bivariate timeseries and each time-segment >

Maximum order of integration (I(d)) – that is, the number of differences required to bring a given timeseries into stationarity – of the full timeseries was determined to be I(1), given that it is the maximum of 0 or 1 with or without considering structural breakpoints (that is, per segment).

## Discussion and Conclusion:

The DIA model offers a formal normative scaffold to explore variables of interest to innovation on the company, sector, industry, or national basis for science and technology firms with a specific focus on securing economic rents from specific forms of IP (notably, patents and trademarks) and their communication and subsequent monetization (Daizadeh, 2007, 2009, 2006, 2007b). While inquiries into the DIA model were restricted to only broad correlation analysis and a specific case study, additional work is needed to further validate the model. Importantly, this additional work may also provide insights into metrics investigating innovative productivities of firms.

Specifically, the DIA model suggests an inter-relationship between various metrics. Here, the inter-relationship between trademark filings (Trademarks) and patent applications (Patents) are explored using a set of statistical analysis that seek to empirically identify structural similarities between the temporal evolution of Trademarks and Patents. The descriptive analysis demonstrated that the distributions of the timeseries are similar. Correlation and cross-wavelet analysis clearly showed synchronicity and coherence between the timeseries. Cointegration analysis demonstrates a 'long-run' equilibrium (restricting divergence) has been established between the timeseries.

The DIA model proposes that R&D expenditure is a driver in patent and trademark originations. This is consistent with the time series (cointegration) analysis performed by Verbeek and Debackere (2006). These authors find "patent evolution is strongly related to … levels of public and private R&D expenditure… (see abstract and conclusions in ibid)."

With regards to the dating of structural breakpoints in the bivariate time series, while additional statistical work is required to better understand sensitivity (as different approaches may realize different dates), it is challenging to link economic shocks that may have caused the abrupt concomitant temporal perturbations to the dates of simultaneous / near simultaneous shocks in the bivariate time series (viz., 1988, 1993/1994, 1999/2000, and 2011). For example, one can hypothesize (and therefore test) that domestic economic hardships affecting R&D (e.g., the dot-com crisis) or the end of a bull market may have been a direct or contributing factor to the abrupt temporal changes in the IP-assets of S&T firms such as Patents and Trademarks (see, e.g., Bleoca, 2014; ). Simultaneously, one can also hypothesize that there was an 'event' associated with the start of the recent "bull market" over the last decade that may have been initiative in the early 2010's (potentially irrespective of or in addition to fluctuations in the legal landscape (notably, from a patent perspective, the 'Leahy-Smith America

Invents Act' that became law in 2011[2])). Further work would need to be done to examine such causal factors during the years identified in this work.

Lastly, from a conceptual perspective, the "long-run" equilibrium of R&D expenditure spill-over effects such as IP-related assets (Trademarks and Patents) and the equilibrium (stationary) processes elucidated between the structural breakpoints (herein called 'regimes' – see Table 4) may be recast along the lines of Schumpeter's theory of business cycles and more generally innovation theory. From this work, while there may be abrupt discontinuations (assumed to be due to non-endogenous / exogeneous factors) within short periods of time (approximately < 2 years), overall economic and innovative progress (as defined by several metrics including those of intellectual property) has continued during the relatively long-time course under-study (monthly intervals of over 40 years). The approach taken herein treats Trademarks and Patents as macro-socio-economic variables averaging across degrees (e.g., radical versus incremental), types (e.g., process versus product) and sectors (e.g., biotech versus manufacturing) of innovativeness. Aligned with Schumpeterian thought around business cycles, and in terms of cointegration of certain macro-socio-economic variables, Konstantakis and Michaelides (2017; page 20) note that "it is exactly upon the existence of this equilibrium relationship that Schumpeterian business cycles were founded, since progressive evolution of innovative activity expressed through technology, leads to the evolution of economic activity as a whole."

Given the Results above, there is a strong and intimate inter-relationship exists between Trademarks and Patents. Beyond supporting the DIA model, this work thus adds to the emerging literature (beginning in part with Daizadeh, 2007) that Trademarks should be of interest as an innovation metric as a unique entity and/or in combination with other such metrics.

[2] https://www.govinfo.gov/app/details/PLAW-112publ29

As with any statistical analysis, there are advantages and disadvantages as well as practical aspects (e.g., computational intensity or algorithm complexity) of the methodologies used within the constraints of the data collected (Daizadeh, 2020). Thus, multiple, complementary, and orthogonal methods to investigate the inter-relationship were used. For example, wavelet analysis is well-known to be of utility across a broad range of implementations (e.g., cross-wavelet) with non-stationary timeseries (Rhif et al, 2019) well complemented both the decomposition (notably seasonal effects) results and the correlation coefficient calculations; two different cointegration tests were performed: Johansen trace test (Johansen, 1988) and Phillips and Ouliaris test (Phillips and Ouliaris, 1990).

The analysis presented in this paper provide supportive evidence for a component of the DIA model as well as metrics tracking innovativeness, however, much further inquiries remain. Future experiments may include:

- Geography: e.g., ex-US versus US inter-relationship inquiries
- Additional trademark and patent variables: e.g., granted patents and designated trademarks
- Integration of additional DIA variables: e.g., numbers of press releases over time and financial metrics
- Deepened analysis: Mapping identified structural breakpoints to the introduction of promulgation of new/updated legal frameworks and/or new technologies to better understand impact
- Comparison of IP-metrics (e.g., trademarkmetrics/patentmetrics) to broader econometrics and scientometrics: e.g., comparing I(1) processes

Figure 1: Timeseries decomposition of Trademarks

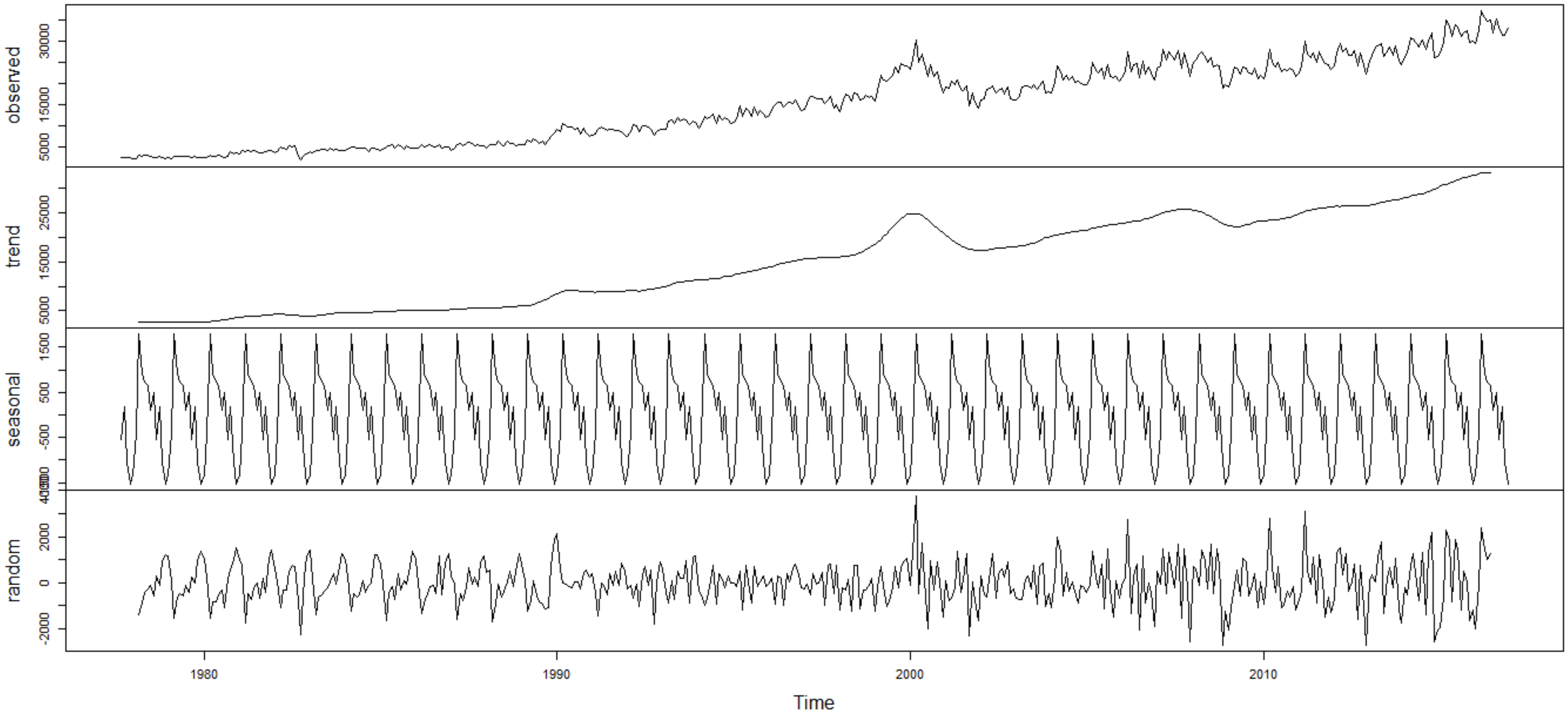

Figure 2: Timeseries decomposition of Patents

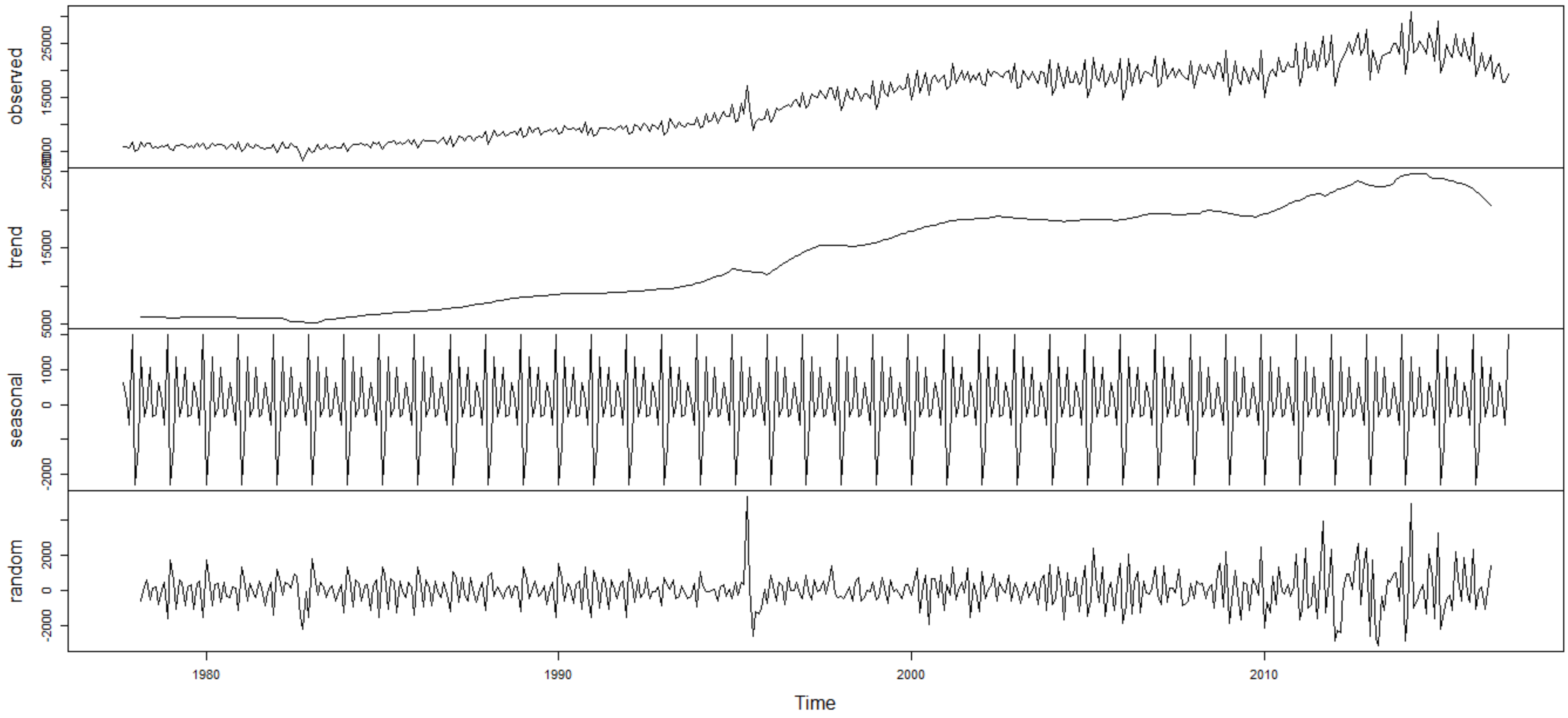

Figure 3: Cross-wavelet analysis of Trademarks and Patents; solid black contour lines designate the 5% significance

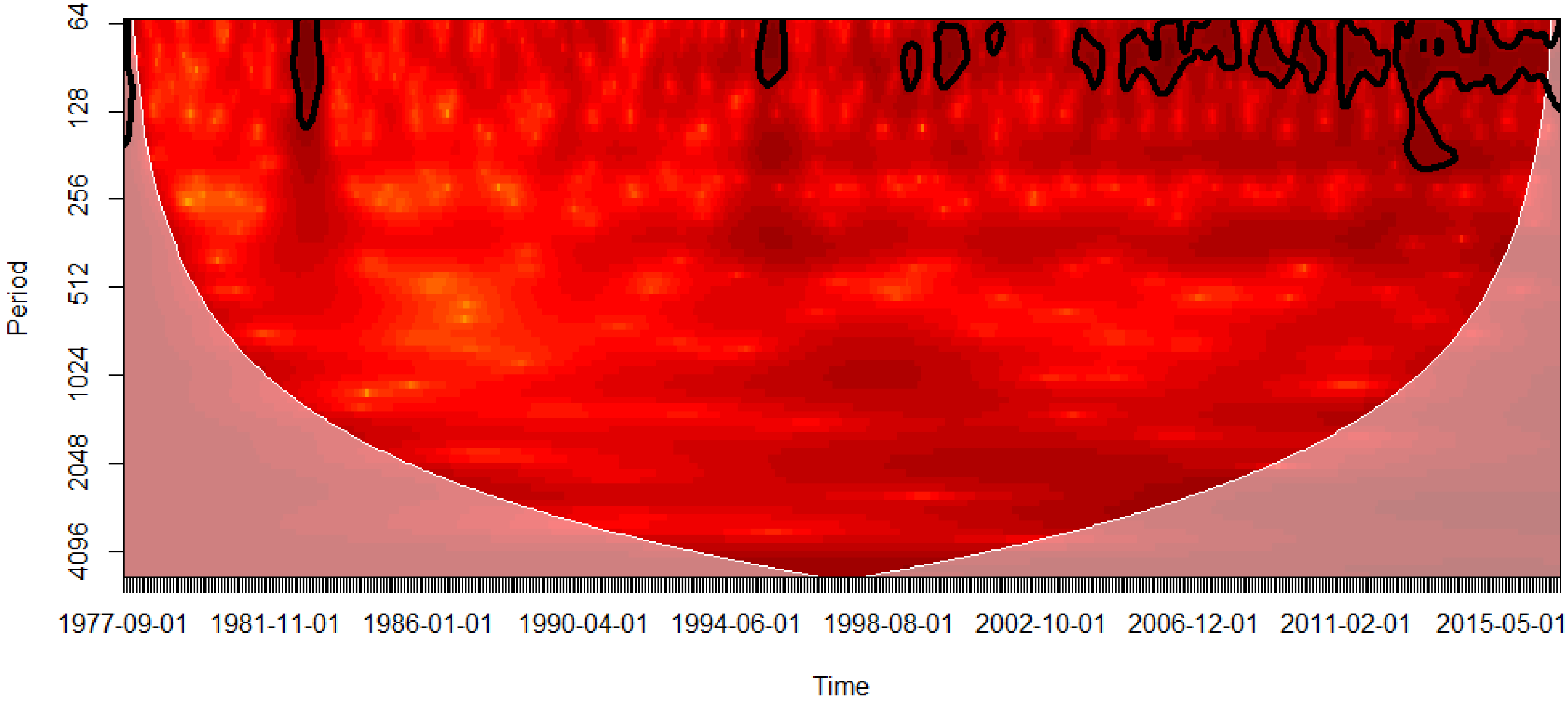

Figure 4: Existence of structural breakpoints for Trademarks

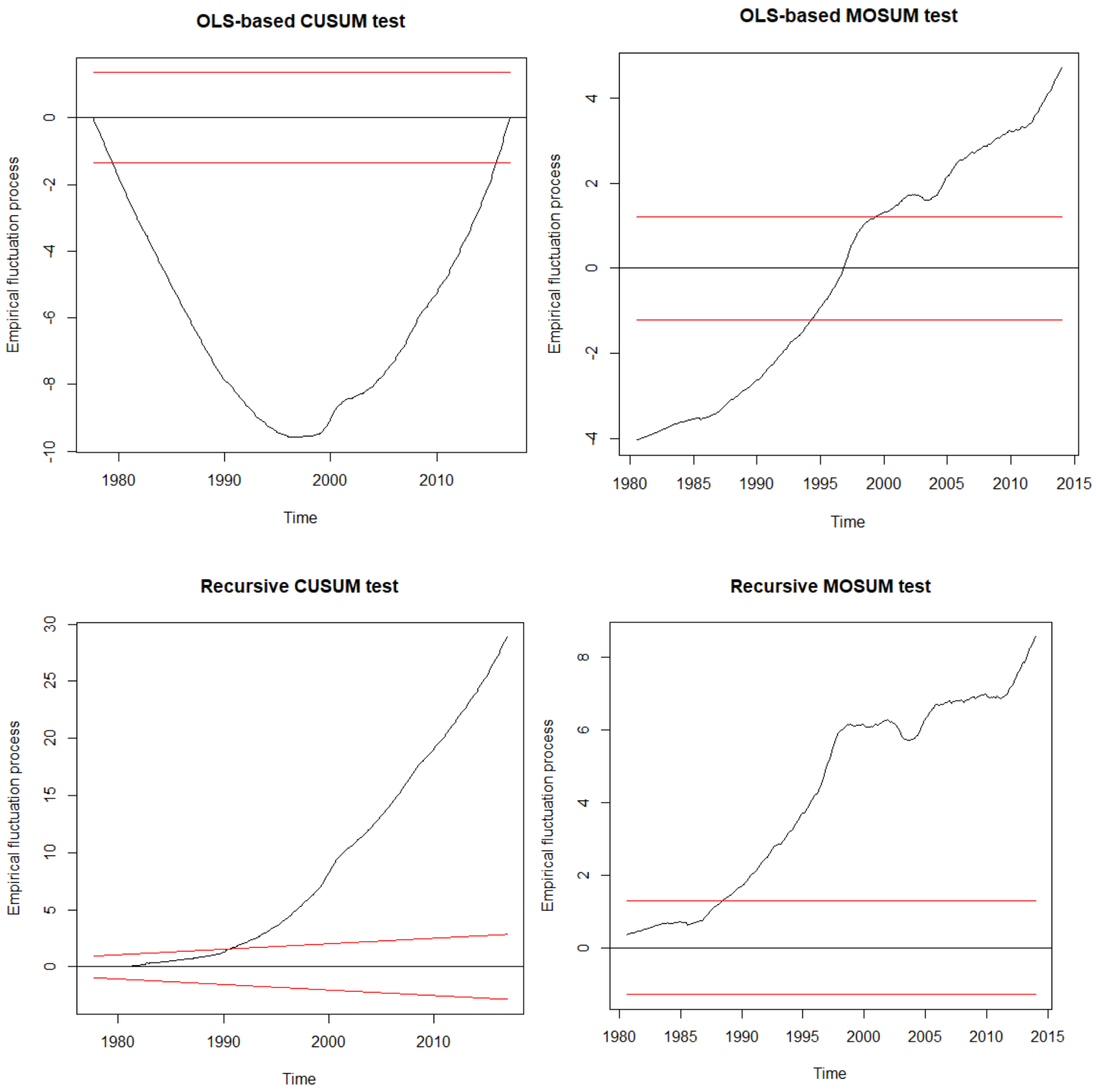

Figure 5: Structural breakpoints with corresponding confidence intervals (see Table 3) identified in Trademarks (black) and Patents (red)

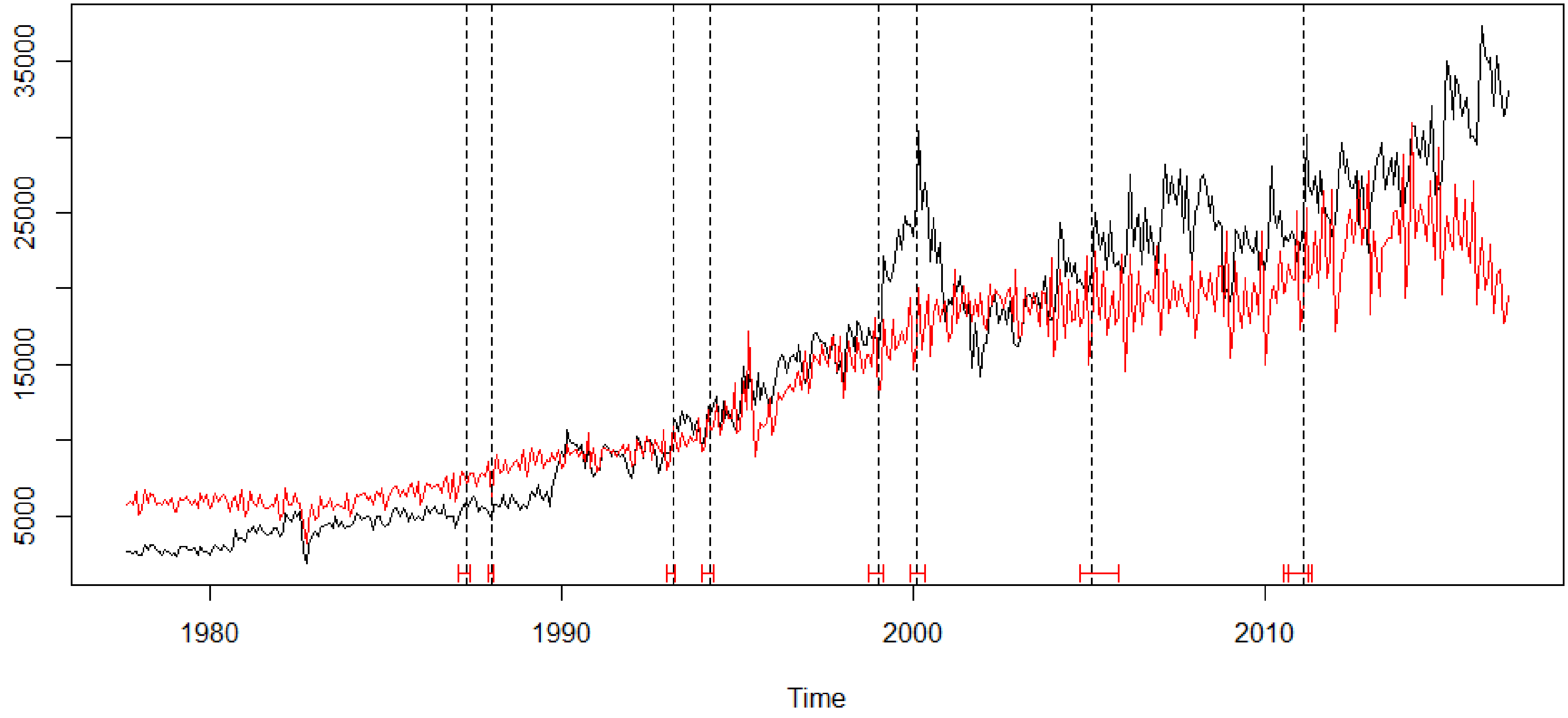

Table 1: Descriptive statistics of Trademarks and Patents

| Variable | Minimum | 1st Quartile | Median | Mean | 3rd Quartile | Maximum | Standard Deviation | Skewness | Kurtosis |
|---|---|---|---|---|---|---|---|---|---|
| Trademarks | 1895 | 5537 | 15456 | 15276 | 23574 | 37317 | 9418.054 | 0.202 | 1.76 |
| Patents | 3134 | 7598 | 14468 | 13930 | 19422 | 30969 | 6523.347 | 0.185 | 1.76 |

Table 2: Significance testing (p-value) for the existence of structural breakpoints in Trademarks and Patents

| Variable | OLS-CUSUM | OLS-MOSUM | REC-CUSUM | REC-MOSUM |
|---|---|---|---|---|
| Trademarks | < 2.2e-16 | 0.01 | < 2.2e-16 | 0.01 |
| Patents | < 2.2e-16 | 0.01 | < 2.2e-16 | 0.01 |

Table 3: Dating (via Bai-Perron) of the structural break points in Trademarks and Patents

| Trademarks | | | Patents | | |
|---|---|---|---|---|---|
| 2.7% | Breakpoint | 97.5% | 2.7% | Breakpoint | 97.5% |
| Feb 1987 | May 1987 | Jul 1987 | Jan 1988 | Feb 1988 | Apr 1988 |
| Jan 1993 | Mar 1993 | Apr 1993 | Dec 1993 | Apr 1994 | May 1994 |
| Oct 1998 | Jan 1999 | Mar 1999 | Dec 1999 | Feb 2000 | Jul 2000 |
| Oct 2004 | Feb 2005 | Nov 2005 | | | |
| Sept 2010 | Feb 2011 | Apr 2011 | Apr 2010 | Feb 2011 | Apr 2011 |

Table 4: Segments identified as longest length of time between Trademarks and Patents structural break points (see Results for description)

| Segment | Heuristic: Longest time between Trademarks and Patents structural break points* |
|---|---|
| 1 | Sep 1977 to April 1987 |
| 2 | Feb 1988 to Feb 1993 |
| 3 | May 1994 to Dec 1998 |
| 4 | Mar 2000 to Jan 2005 |
| 5 | Mar 2005 to Jan 2011 |
| 6 | Mar 2011 to Dec 2016 |
| *That is, the origin to the month prior to breakpoint: October 1983 | |

Table 5: Results of cointegration tests across full bivariate timeseries and each time-segment

| Segment | Johansen procedure (trace statistics, without linear trend and constant) | | Phillips and Ouliaris Test |
|---|---|---|---|
| | r<=1: Test statistic versus (v) critical value of test (at lowest level of alpha) | r=0 Test statistic versus (v) critical value of test (at lowest level of alpha) | Value of Test Statistic versus (v) critical value of test (at lowest level of alpha) |
| Full timeseries | [redacted] | 75.47 v 24.60 (1%) | 222.6575 v 55.1911 (1%) |
| 1 | [redacted] | 23.37 v 19.96 (5%) | [redacted] |
| 2 | [redacted] | 36.34 v 24.60 (1%) | [redacted] |
| 3 | [redacted] | 24.64 v 24.60 (1%) | 51.1519 v 40.8217 (5%) |
| 4 | 9.95 v 9.24 (5%) | 74.61 v 24.60 (1%) | [redacted] |
| 5 | 11.58 v 9.24 (5%) | 83.56 v 24.60 (1%) | 83.8507 v 55.1911 (1%) |
| 6 | [redacted] | 33.83 v 24.60 (1%) | 54.1304 v 40.8217 (5%) |

Table 6: Number of differences required to bring the time-series into stationarity.

| Variable | Segment | KPSS | ADF | PP |
|---|---|---|---|---|
| Trademarks | Full dataset | 1 | 1 | 1 |
| Patents | | 1 | 1 | 1 |
| Trademarks | 1 | 1 | 1 | 1 |
| Patents | | 1 | 0 | 0 |
| Trademarks | 2 | 1 | 1 | 1 |
| Patents | | 1 | 0 | 0 |
| Trademarks | 3 | 1 | 1 | 1 |
| Patents | | 1 | 1 | 0 |
| Trademarks | 4 | 1 | 0 | 0 |
| Patents | | 0 | 0 | 0 |
| Trademarks | 5 | 0 | 0 | 0 |
| Patents | | 0 | 0 | 0 |
| Trademarks | 6 | 1 | 1 | 1 |
| Patents | | 0 | 0 | 0 |

#### - Begin Supplementary Materials ####

#Trademarks
#Go to TESS: http://tmsearch.uspto.gov/bin/gate.exe?f=tess&state=4804:57thz4.1.1
# Manually search and collect Number of Trademarks as follows:
#By Filing Date: "(198712$)[FD]" - Where 198712$ is the %Y%m$.

#Patents
#Go to PATFT: http://patft.uspto.gov/netahtml/PTO/search-adv.htm
#By Application Filing Date: "APD/12/$/2018"

#The patent and trademark filings data were collected from Sept 1977 to Dec 2018.

#Confirm version of R:

```
> citation()
```

R Core Team (2019). R: A language and environment for statistical computing. R Foundation for Statistical Computing, Vienna, Austria. URL https://www.R-project.org/.

```
> version
_
platform       x86_64-w64-mingw32
arch           x86_64
os             mingw32
system         x86_64, mingw32
status
major          3
minor          6.1
year           2019
month          07
day            05
svn rev        76782
language       R
version.string R version 3.6.1 (2019-07-05)
nickname       Action of the Toes
```

#Read into R:

```
> IP<- read.csv("C:/Users/DAIZAI/Desktop/patent/Patent-Trademark.csv", sep=",")
```

#Confirm dataframe - length/variables and shrink by 24m to avoid so-called 'patent cliff':

```
> str(IP)
```

```
'data.frame':   496 obs. of  3 variables:
$ Date: Factor w/ 496 levels "1/1/1978","1/1/1979",..: 455 42 84 126 1 168 209 250 291 332 ...
$ Number.of.Trademark.Applications: int  2669 2597 2552 2604 2386 2370 3126 2738 3028 3088 ...
$ Number.of.Patent.Applications   : int  5760 5898 5731 6630 5064 5439 6660 5799 6487 6419 ...

# Shrinking by 24 months dues to so-called "patent-cliff"
> TrademarksTotal<-IP$Number.of.Trademark.Applications[1:472]
> PatentsTotal<-IP$Number.of.Patent.Applications[1:472]

 #Convert to Time-Series, decompose time-series, and perform descriptive statistics
> tsTrademarks<-ts(TrademarksTotal,start=c(1977,9),frequency=12)
> tsPatents<-ts(PatentsTotal,start=c(1977,9),frequency=12)

> tsTrademarks

       Jan   Feb   Mar   Apr   May   Jun   Jul   Aug   Sep   Oct   Nov   Dec

1977                                          2669  2597  2552  2604
1978  2386  2370  3126  2738  3028  3088  2708  2638  2465  2793  2636  2362
1979  2518  2350  2920  2968  2953  2794  2741  2829  2438  2956  2676  2596
1980  2469  2607  3035  2893  2797  3094  2883  2590  2928  4081  3412  3559
1981  3329  4113  3906  4297  3871  4358  4077  3815  3688  3879  4162  4140
1982  3594  4009  5128  4868  4576  5244  4942  5264 15843  1895  3126  3529
1983  3915  3597  4224  4297  4389  4543  4192  4893  4265  4634  4189  4296
1984  4281  4472  5167  5068  4762  4837  4914  4803  4064  4896  4922  4474
1985  4296  4408  5087  5417  5537  4914  5537  5215  4627  5218  4747  4904
1986  4785  4677  5569  5153  5397  5630  5152  5124  5715  4787  4950  4931
1987  4188  4826  5528  5780  5372  5860  6269  5990  5309  5624  5454  5286
1988  4758  5756  5730  5618  6358  5706  5473  6341  5776  5899  5384  5744
1989  5546  5761  6615  6230  7071  6496  5995  6567  5613  6687 11400  8450
1990  9209  8797 10687  9936  9836  9707  9319  9634  8167  9567  8417  7546
1991  7906  7986  9192  9734  9441  8962  9371  9260  8884  9073  8805  7972
1992  7486  8612 10248  9940  8776 10078 10019  9701  9348  7853  8599  9107
1993  9143  9143 11059 11352 10594 11897 11160 11628 11448 10374 11248 10923
1994  9599 10388 12173 11664 12292 12902 10728 12536 11443 11932 11339 10634
1995 10961 11857 14860 12391 14323 13644 12325 14434 12631 13726 12926 12072
1996 12582 13872 15117 15636 15565 14433 15447 15590 15234 16299 14727 13810
1997 14060 15229 16962 17079 16460 16465 16425 15363 16089 16901 14367 15163
1998 13510 15465 17624 17224 15805 17862 17515 16270 16331 17426 16737 16897
1999 16002 18431 22143 20957 20515 24106 21999 23891 22563 24753 24225 24272
2000 23412 25815 30423 25204 26986 24831 21769 24626 21644 22962 20340 17822
2001 19248 18923 20732 19781 20778 18551 18311 19608 14770 17727 15587 14181
2002 16273 16357 18555 18946 19622 17734 18359 18958 17639 19111 16487 16278
```

```
2003 16170 17077 19352 19461 19563 19045 19687 18614 19518 20752 17857 18083
2004 17848 20221 24380 22857 20851 21971 20794 21686 20329 20545 20163 19727
2005 19852 21156 25021 23106 22583 23621 21283 24405 21769 21419 21833 20716
2006 21532 22810 27578 22750 24386 24861 21580 25319 22311 24077 22658 20909
2007 23774 24365 28236 25640 27467 26565 25520 27860 23629 27407 24521 21614
2008 25011 25689 27048 27577 26586 24960 26086 23857 24425 24124 19074 19873
2009 19111 20849 23869 23573 22337 24084 23578 22625 22336 23795 21248 21987
2010 21221 23097 28105 25119 23953 25080 22796 23413 23092 23739 23151 21794
2011 22686 24232 30145 26896 26187 27407 25024 27708 25636 24807 24550 23480
2012 24295 27305 29660 27660 28521 26754 26636 27794 24270 27369 24447 22310
2013 25382 26630 28633 29002 29609 26566 27697 28612 26811 28991 25622 24625
2014 26543 27243 30853 30752 29450 28680 30393 28186 30059 32019 26212 26558
2015 27182 29681 35025 33786 31123 34043 33316 31327 32138 32584 29909 30094
2016 29461 32456 37317 35506 34861 35211 32027 35356 33293 31394 31452 33063
```

```
> tsPatents
       Jan   Feb   Mar   Apr   May   Jun   Jul   Aug   Sep   Oct   Nov   Dec

1977                               5760  5898  5731  6630
1978  5064  5439  6660  5799  6487  6419  5671  5831  5697  6012  5756  6210
1979  5303  5240  6131  6071  6247  6087  5726  5999  5541  6459  5913  6399
1980  5492  5619  6491  5971  6292  6325  6077  5328  6143  6246  5414  6726
1981  4899  5351  6548  5857  5583  6377  5907  5539  5489  5726  5602  6354
1982  4677  5313  6793  5724  5702  6543  5894  5834 10870  3134  4604  5692
1983  4698  4900  6279  5451  5657  6202  5390  5621  5740  5648  5544  6507
1984  4940  5557  6360  6216  6391  6522  6033  6230  5708  6631  6319  6774
1985  5381  6111  6664  6807  6895  6257  6713  6290  6814  7204  6247  7251
1986  5730  6296  7137  7026  6837  6982  6994  6486  7150  7544  6200  7814
1987  5918  6753  7899  7633  7027  7852  7784  6968  7510  7814  7539  8614
1988  6315  7682  8992  7850  8040  8737  7843  8262  8551  8721  8082  9321
1989  7616  8117  9466  8379  9123  9378  8133  8702  8588  9083  8706  9308
1990  8151  8472  9707  9061  9197  9331  8963  9269  8533 10444  8128  9421
1991  7879  8270  9363  9413  9406  9080  9394  8954  9327  9737  9155  9704
1992  8237  8485 10025  9591  8881 10282  9822  8666 10009  9586  9180 10693
1993  8063  8728 11064  9963  9246 10500  9793  9531 10232  9904 10034 11401
1994  9243  9614 12007 10453 10893 12276 10303 11300 12499 11444 11444 13760
1995 10445 10716 13835 11945 17222 28123  8860 10340 11124 10913 11152 12746
1996 10346 11112 13067 12689 13000 13285 13677 13195 14326 14524 13319 15858
1997 13121 13531 15641 15287 15015 16286 15456 14815 16658 16871 14529 16885
1998 12722 13746 16543 15062 14511 16769 15660 14426 15607 15603 14861 18092
1999 12865 14204 17997 15791 15337 17893 15963 16308 17160 16447 16702 19417
2000 14586 16451 20021 15922 17884 19641 15521 18278 19273 17834 18532 19122
```

```
2001 16493 17127 21285 17692 18593 20122 18202 19753 17856 19239 18210 19660
2002 17694 17281 20278 18963 19891 19568 19240 18754 19562 20034 17799 21302
2003 16740 16898 19990 18907 18242 19748 18830 17542 19825 19708 16896 22063
2004 15528 16931 21274 18212 16712 20593 17779 17943 19884 17521 18130 22102
2005 14968 17021 22490 18546 17987 21090 16914 18462 19787 17592 18133 22188
2006 14550 16701 22216 17208 19402 21150 17560 19571 19828 19312 18999 22766
2007 16930 17539 22271 18427 19480 20506 18797 20267 19143 25045 18195 21820
2008 16717 18825 21168 19762 19318 21035 19896 18546 21495 21413 18120 23736
2009 15375 17696 21848 18816 17438 20736 19164 17736 20407 19603 18327 23801
2010 14972 17696 19786 20199 19019 22414 19760 19867 21604 20657 20612 25143
2011 17261 18992 25367 20428 20964 23822 20058 22268 26472 20658 21926 26522
2012 17222 19167 21736 22546 23921 25237 23054 24756 27051 22929 24437 27708
2013 18316 23728 42788 19501 22634 22932 23350 23313 25042 25126 23015 28838
2014 19391 22253 30969 23387 24040 25581 24700 23086 27091 25141 21855 29294
2015 19599 21643 24767 23184 22546 26852 23783 22590 25856 23159 21726 27049
2016 18970 20933 23298 19948 20837 22912 18417 20752 21299 17774 17791 19436
```

```
> plot(decompose(tsTrademarks,type="additive"))
> plot(decompose(tsPatents,type="additive"))
```

```
#Identify outliers
#Javier López-de-Lacalle (2019). tsoutliers: Detection of Outliers in Time Series. R package version 0.6-8.
# https://CRAN.R-project.org/package=tsoutliers
library(tsoutliers)
> TrademarksOutliers<-tso(tsTrademarks,types = c("AO","LS","TC"),maxit.iloop=10)
> PatentsOutliers<-tso(tsPatents,types = c("AO","LS","TC"),maxit.iloop=10)
```

```
> TrademarksOutliers
Series: tsTrademarks
Regression with ARIMA(2,1,1)(0,1,2)[12] errors
Coefficients:
        ar1      ar2     ma1     sma1     sma2        AO61      LS147      LS262
-1.0107  -0.5826  0.4306  -0.4939  -0.2779  12137.5320  4527.2969  3409.3950
s.e.   0.0834   0.0440  0.1067   0.0480   0.0458    669.2913   681.1637   674.9746
sigma^2 estimated as 868751:  log likelihood=-3790.79
AIC=7599.58   AICc=7599.98   BIC=7636.74
```

```
Outliers:
  type ind    time coefhat  tstat
1   AO  61 1982:09   12138 18.135
2   LS 147 1989:11    4527  6.646
3   LS 262 1999:06    3409  5.051
```

```
> PatentsOutliers
Series: tsPatents
Regression with ARIMA(3,0,0)(2,1,2)[12] errors
Coefficients:
        ar1     ar2     ar3    sar1     sar2     sma1    sma2       AO61       AO214
     0.2731  0.2776  0.4185  0.7318  -0.3230  -1.4584  0.6303  5591.8986  15515.5416
s.e.  0.0480  0.0438  0.0468  0.1133   0.0673   0.1146  0.0879   773.2661    764.2898
         AO362      AO427
     5058.3040  17555.5353
s.e.   757.5541    799.8064
sigma^2 estimated as 917822:  log likelihood=-3812.79
AIC=7649.58   AICc=7650.27   BIC=7699.15

Outliers:
  type ind    time coefhat  tstat
1   AO  61 1982:09    5592  7.232
2   AO 214 1995:06   15516 20.301
3   AO 362 2007:10    5058  6.677
4   AO 427 2013:03   17556 21.950

> plot(TrademarksOutliers); X11(); plot(PatentsOutliers)

#Clean/smooth data - replace identified outliers (X) with average of prior (X(t-1)) and posterier (X(t+1))

>Trademarks<-tsTrademarks; Patents<-tsPatents
>Trademarks[61]= (Trademarks[62]+Trademarks[64]) / 2
>Trademarks[147]= (Trademarks[146]+Trademarks[148]) / 2
>Trademarks[262]= (Trademarks[261]+Trademarks[263]) / 2
>Patents[61]= (Patents[62]+Patents[64]) / 2
>Patents[214]= (Patents[213]+Patents[215]) / 2
>Patents[362]= (Patents[361]+Patents[363]) / 2
>Patents[427]= (Patents[426]+Patents[428]) / 2

> plot(decompose(Trademarks,type="additive"))
> plot(decompose(Patents,type="additive"))

> library(moments); citation("moments")

#Lukasz Komsta and Frederick Novomestky (2015). moments: Moments, cumulants, skewness, kurtosis
#and related tests. R package version 0.14. https://CRAN.R-project.org/package=moments

#Use fitted output from tsoutliers

>summary(Trademarks); sd(Trademarks); skewness(Trademarks); kurtosis(Trademarks)
>summary(Patents); sd(Patents); skewness(Patents); kurtosis(Patents)
```

```
#note: skew/kurtosis comparative - no need to transform

#Perform correlation analysis, auto- and partial-correlation and cross-wavelet
>cor(Trademarks, Patents, method=”spearman”);
[1] 0.9431803
>cor(Trademarks, Patents, method=”kendall”);
[1] 0.8024742

#Perform cross-wavelet analysis
> DATE<-seq(as.Date("1977/9/01"), as.Date("2016/12/01"), "months")
> tTrademarks <- cbind(DATE, Trademarks)
> tPatents <- cbind(DATE, Patents)
> XWTradePatent<-xwt(tTrademarks,tPatents)
Warning messages:
1: In arima(d1[, 2], order = c(1, 0, 0), method = arima.method): possible convergence problem: optim
gave code = 1
2: In arima(x, order = c(1, 0, 0), method = arima.method): possible convergence problem: optim gave
code = 1
> plot(XWTradePatent, xaxt=”n”)
> axis(side=1, at=c(seq(as.Date("1977/9/01"), as.Date("2016/12/01"), "months")),
labels=c(seq(as.Date("1977/9/01"), as.Date("2016/12/01"), "months")))

#Perform Structural Change Analysis: Confirm existence of structural break within timeseries using
#Achim Zeileis, Friedrich Leisch, Kurt Hornik and Christian Kleiber (2002). strucchange: An R Package for
#Testing for Structural Change in Linear Regression Models. Journal of Statistical Software, 7(2), 1-38.
#URL http://www.jstatsoft.org/v07/i02/

#Achim Zeileis, Christian Kleiber, Walter Kraemer and Kurt Hornik (2003). Testing and Dating o
#Structural Changes in Practice. Computational Statistics & Data Analysis, 44, 109-123.
>library(strucchange)

#Trademarks: OLS-CUSUM/OLS-MOSUM/#REC-CUSUM/REC-MOSUM
>Trademarks.olscus<- efp(Trademarks~1, type="OLS-CUSUM"); plot(Trademarks.olscus)
>Trademarks.olsmus<- efp(Trademarks~1, type="OLS-MOSUM"); plot(Trademarks.olsmus)
>Trademarks.reccus<- efp(Trademarks~1, type="Rec-CUSUM"); plot(Trademarks.reccus)
>Trademarks.recmus<- efp(Trademarks~1, type="Rec-MOSUM"); plot(Trademarks.recmus)

#Patents: OLS-CUSUM/OLS-MOSUM/#REC-CUSUM/REC-MOSUM
>Patents.olscus<- efp(Patents~1, type="OLS-CUSUM"); plot(Patents.olscus)
>Patents.olsmus<- efp(Patents~1, type="OLS-MOSUM"); plot(Patents.olsmus)
>Patents.reccus<- efp(Patents~1, type="Rec-CUSUM"); plot(Patents.reccus)
>Patents.recmus<- efp(Patents~1, type="Rec-MOSUM"); plot(Patents.recmus)

#Perform significance tests for Empirical fluctuation processes: Null hypothesis: No structural change
```

```
>sctest(Trademarks.olscus);sctest(Trademarks.olsmus);sctest(Trademarks.reccus);sctest(Trademarks.recmus)
>sctest(Patents.olscus);sctest(Patents.olsmus);sctest(Patents.reccus);sctest(Patents.recmus)

#Perform dating of structural change / break points via Bai-Perron

#Tradenames:
>bTrademarks<-breakpoints(Trademarks~1)
>cTrademarks<-confint(bTrademarks)

#Patents:
>bPatents<-breakpoints(Patents~1)
>cPatents<-confint(bPatents)

>cTrademarks; cPatents

>library (tseries); seqplot.ts(Trademarks,Patents); lines(cTrademarks); lines(cPatents) #R package tseries
#(Trapletti and Hornik, 2019)

> bTrademarks; bPatents

#Based on SBPs, determine segments through heuristic of longest time between any two SBPs:
#Segment 1: 1:116              Sep 1977 to April 1987
#Segment 2: 126:186            Feb 1988 to Feb 1993
#Segment 3: 201:256            May 1994 to Dec 1998
#Segment 4: 271:329            Mar 2000 to Jan 2005
#Segment 5: 331:401            Mar 2005 to Jan 2011
#Segment 6: 403:472            Mar 2011 to Dec 2016

> tseg1<-Trademarks[1:116]; tseg2<-Trademarks[126:186]; tseg3<-Trademarks[201:256]; tseg4<-Trademarks[271:329]; tseg5<-Trademarks[331:401]; tseg6<-Trademarks[403:472]

> pseg1<-Patents[1:116]; pseg2<-Patents[126:186]; pseg3<-Patents[201:256]; pseg4<-Patents[271:329]; pseg5<-Patents[331:401]; pseg6<-Patents[403:472]

> Segment0<- as.matrix(as.data.frame(cbind(Trademarks, Patents))) #Full Dataset
> Segment1<- as.matrix(as.data.frame(cbind(Trademarks[1:116],Patents[1:116])))
> Segment2<- as.matrix(as.data.frame(cbind(Trademarks[126:186],Patents[126:186])))
> Segment3<- as.matrix(as.data.frame(cbind(Trademarks[201:256],Patents[201:256])))
> Segment4<- as.matrix(as.data.frame(cbind(Trademarks[271:329],Patents[271:329])))
> Segment5<- as.matrix(as.data.frame(cbind(Trademarks[331:401],Patents[331:401])))
> Segment6<- as.matrix(as.data.frame(cbind(Trademarks[403:472],Patents[403:472])))

#Cointegration analyses: Test for cointegration across entire and then sections
```

#Adrian Trapletti and Kurt Hornik (2019). tseries: Time Series Analysis and Computational Finance. R
#package version 0.10-47.
#Pfaff, B. (2008) Analysis of Integrated and Cointegrated Time Series with R. Second Edition. Springer,
#New York. ISBN 0-387-27960-1

```
> install.packages("tseries"); library(tseries)
> install.packages("urca"); library(urca)
```

#Johansen Procedure

```
> summary(ca.jo(Segment0, ecdet="const",type="trace"))
> summary(ca.jo(Segment1, ecdet="const",type="trace"))
> summary(ca.jo(Segment2, ecdet="const",type="trace"))
> summary(ca.jo(Segment3, ecdet="const",type="trace"))
> summary(ca.jo(Segment4, ecdet="const",type="trace"))
> summary(ca.jo(Segment5, ecdet="const",type="trace"))
> summary(ca.jo(Segment6, ecdet="const",type="trace"))
```

Philips and Ouliaris Test

```
>summary(ca.po(Segment0, type= "Pz"))
>summary(ca.po(Segment1, type= "Pz"))
>summary(ca.po(Segment2, type= "Pz"))
>summary(ca.po(Segment3, type= "Pz"))
>summary(ca.po(Segment4, type= "Pz"))
>summary(ca.po(Segment5, type= "Pz"))
>summary(ca.po(Segment6, type= "Pz"))
```

#Perform unit and stationarity assessments

```
> library(forecast)
>ndiffs(Trademarks, test="kpss"); ndiffs(Trademarks, test="adf"); ndiffs(Trademarks, test="pp")
>ndiffs(Patents, test="kpss"); ndiffs(Patents, test="adf"); ndiffs(Patents, test="pp")

>ndiffs(tseg1, test="kpss"); ndiffs(tseg1, test="adf"); ndiffs(tseg1, test="pp")
>ndiffs(pseg1, test="kpss"); ndiffs(pseg1, test="adf"); ndiffs(pseg1, test="pp")

>ndiffs(tseg2, test="kpss"); ndiffs(tseg2, test="adf"); ndiffs(tseg2, test="pp")
>ndiffs(pseg2, test="kpss"); ndiffs(pseg2, test="adf"); ndiffs(pseg2, test="pp")

>ndiffs(tseg3, test="kpss"); ndiffs(tseg3, test="adf"); ndiffs(tseg3, test="pp")
>ndiffs(pseg3, test="kpss"); ndiffs(pseg3, test="adf"); ndiffs(pseg3, test="pp")

>ndiffs(tseg4, test="kpss"); ndiffs(tseg4, test="adf"); ndiffs(tseg4, test="pp")
>ndiffs(pseg4, test="kpss"); ndiffs(pseg4, test="adf"); ndiffs(pseg4, test="pp")

>ndiffs(tseg5, test="kpss"); ndiffs(tseg5, test="adf"); ndiffs(tseg5, test="pp")
>ndiffs(pseg5, test="kpss"); ndiffs(pseg5, test="adf"); ndiffs(pseg5, test="pp")
```

```
>ndiffs(tseg6, test=”kpss”); ndiffs(tseg6, test=”adf”); ndiffs(tseg6, test=”pp”)
>ndiffs(pseg6, test=”kpss”); ndiffs(pseg6, test=”adf”); ndiffs(pseg6, test=”pp”)
```

#### - End Supplementary Materials ####